# HIGH POWER TEST OF A 3.9 GHZ 5-CELL DEFLECTING-MODE CAVITY IN A CRYOGENIC OPERATION

Young-Min Shin[1, 2], and Michael Church[2]
[1]Department of Physics, Northern Illinois University, Dekalb, IL, 60115, USA
[2]Fermi National Accelerator Laboratory (FNAL), Batavia, IL 60510, USA

*Abstract*

A 3.9 GHz deflecting mode ($\pi$, $TM_{110}$) cavity has been long used for six-dimensional phase-space beam manipulation tests [1-5] at the A0 Photo-Injector Lab (16 MeV) in Fermilab and their extended applications with vacuum cryomodules are currently planned at the Advanced Superconducting Test Accelerator (ASTA) user facility (> 50 MeV). Despite the successful test results, the cavity, however, demonstrated limited RF performance during liquid nitrogen ($LN_2$) ambient operation that was inferior to theoretical prediction. We have been performing full analysis of the designed cavity by analytic calculation and comprehensive system simulation analysis to solve complex thermodynamics and mechanical stresses. The re-assembled cryomodule is currently under the test with a 50 kW klystron at the Fermilab A0 beamline, which will benchmark the modeling analysis. The test result will be used to design vacuum cryomodules for the 3.9 GHz deflecting mode cavity that will be employed at the ASTA facility for beam diagnostics and phase-space control.

## INTRODUCTION

Over the past decade, a multi-cell deflecting ($TM_{110}$) mode cavity has been employed for phase-space manipulation tests of high brightness beams [1-6] at the Fermilab A0 photoinjector (A0PI), and extended applications are currently scheduled at the ASTA user facility (> 50 MeV). Despite the past successful experimental results, the cavity demonstrated a limited RF performance during liquid nitrogen ($LN_2$) operation, which did not reach the theoretically predicted gradient. The designed cavity has been fully examined with theoretical calculations, based on the Panofsky-Wenzel theorem, using an integrated modeling tool with a comprehensive system analysis capacity to solve complex thermodynamics and the mechanical stress of the multi-cell. This paper discusses the cryogenic RF performance of the 5-cell deflecting mode cavity with numerical modeling analysis. It also presents up-to-date test simulation results of an integrated thermo-stress analysis modeling tool on the deflecting cavity vacuum-cryomodule and low power RF-test results of warm (room-temp, 297 K) and cold ($LN_2$-temp, 80K) cavities.

## SYSTEM DESCRIPTION

At the Fermilab A0PI, the deflecting mode cavity has been used for various beam optics experiments. The cavity was designed with 5 cells to maximize kick strength and powered with a 50 kW (peak), S-band (3.9 GHz) klystron. The RF power was coupled into the cavity through the high power TEM-mode coaxial coupler that was built in the liquid nitrogen ($LN_2$) vessel. The coupler design includes a temperature gradient from cryogenic temperature (80K) of a $LN_2$-ambient cavity to room temperature of an input waveguide. As the emittance exchange only requires modest fields and short pulse lengths, the $TM_{110}$ mode cavity was constructed out of oxygen-free, high conductivity (OFHC) copper [7]. A higher $Q_0$ was required than what was achievable at room temperature with the OFHC copper. We see that $Q_0$ is proportional to the square root of the copper's bulk conductivity. A $Q_0$ 2.4 times greater was achieved by simply incorporating a $LN_2$ cryogenic jacket into the design.

The system is designed with the $LN_2$ vessel because the conductivity of normal conducting copper is increased 6 times from $5.8 \times 10^7$ $\Omega^{-1}m^{-1}$ at room temperature to $3.5 \times 10^8$ $\Omega^{-1}m^{-1}$ at 80 K, which doubles the cavity Q. The cryo-vessel was designed with three frequency tuning screws, attached to the chamber-outside body at one end and the flange, brazed with the beam pipe at the other end, across the flexible bellows. The tuners push the flange against the body and the mechanical pressure is transferred to the cavity through the beam pipe so the structural distortion by the tuner induces frequency change. As this simple design did not include vacuum insulation, the $LN_2$-temperature (~ 80 K) was maintained by shielding the outer body with foam insulation. The coaxial input coupler was designed with the critical matching condition, $\beta = Q_0/Q_e \sim 1$, for maximum RF power coupling into the cavity. The original design includes many practical considerations in various technical aspects. However, a high-Q cavity sensitively responds to dimensional deviations and external perturbations, which could significantly limit the deflecting performance in a cryostat. In particular, structural variation of the input coupler can significantly influence RF coupling characteristics producing an unevenly distributed field profile. It is thus highly probable that the limited deflecting performance of the 5 cell can be attributed to an off-resonance RF coupling presumably owing to design error, fabrication error, and/or cryo-cooling contraction. In order to completely identify the operational constraints, we thus investigated the cavity design and estimated its performance with a theoretical assessment incorporated with numerical data of RF simulation modeling analysis.

We conclude that the off-resonance coupling leads to a reduction in the kick strength because it more likely perturbs the field distribution, decreasing cell-to-cell field uniformity, rather than directly weakening the field strength. The accelerating potential vs. radial position and the deflecting force versus the field deviation. Note that the increasing rate of the deflecting field amplitude along the radial distance is noticeably reduced with loss of field uniformity, which is reflected in the deflecting force: kick strength drops down to 2.9 from 3.95 with 40 % non-uniformity (35.6 % reduction). Eventually, one can see that poorly coupled RF power, accompanying perturbation of the field distribution, possibly causes a 30 ~ 40 % deficiency of the deflecting strength.

## SIMULATION MODELING

The frequency shift ($TM_{110,\pi}$ mode) due to the effective temperature declination from room temperature (300 K) to $LN_2$ (80 K) is explained by the uniform change of the cavity volume. The frequency is increased from 3.8725 GHz at 300 K to 3.8870 GHz at 80 K, which corresponds to 68 kHz/K. The simulation analysis estimates that cryo-cooling causes roughly a 24 MHz frequency deviation to the design cavity that would need to be considered in the process of determining cavity dimensions. This approximate analytic assessment is effective for quantifying an average deviation with the assumption that the entire cavity uniformly contracts with a decrease in temperature. However, it does not accurately predict local displacements due to RF-loading impacts over the structure, in particular on the system with fixed points. Moreover, it is limited in accurately implementing localized effects of critical heat sources such as RF loading and beam loading in the cryomodule analysis. Therefore, we extensively investigated RF-thermal characteristics of this 5-cell cryomodule with full 3D thermal and mechanical simulation modeling analysis. More details are discussed in the next section.

Figure 1 depicts the recently designed vacuum-cryomodule containing the 3.9 GHz 5-cell deflecting mode cavity that could be installed at the Fermilab ASTA user's facility. It is designed with the vacuum insulator, instead of the foam insulator. Volumetric contraction and thermal fluctuation resulting from $LN_2$ cooling is no longer a critical factor since the vacuum insulator tank is an excellent heat reservoir for the $LN_2$ vessel, which linearly cools the ambient temperature down to 77 – 80 K and maintains it consistently. Instead, it is more critical to consider RF loading and beam loading for CW or high duty operation of high intensity or high power machines. The frequency tuner is similar to the original design except that it is mounted on the vacuum tank with an additional flexible bellows. In this design, the amount of liquid nitrogen is controlled by the $N_2$ inlet and vent/relief and its level is monitored by the $LN_2$ level probe. The temperature of the $LN_2$ vessel remains constant by means of the vacuum insulation tank. However, even in the thermally insulated vacuum system energy of long pulse duration deposits a considerable amount of power on the cavity surfaces that can increase the ambient temperature so highly as to exceed the temperature threshold of bubble formation in the $N_2$ fluid.

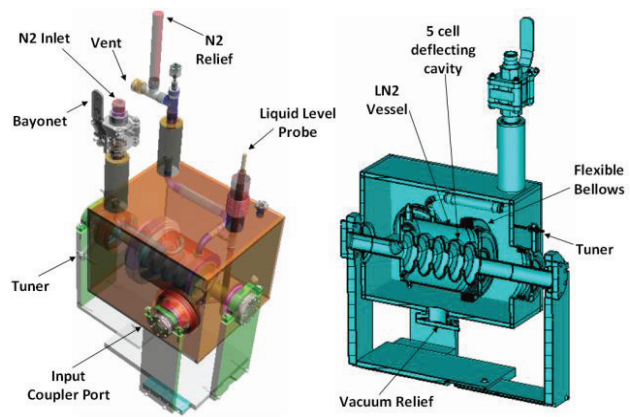

Figure 1: Designed vacuum-cryostat with a 5 cell deflecting mode cavity (a) x-ray image and (b) cross-sectional view

If a deposited power density on the cavity surface exceeds 1 $W/cm^2$ on average, it is highly probable that the ambient temperature exceeds the $LN_2$ boiling point (~ 77 – 80 K) so as to rapidly increase gap pressure and create $N_2$ bubbles that may change the ambient pressure around the cavity in the $LN_2$ vessel. The excessive fluctuation of vessel pressure could substantially impact cavity performance via frequency change, structural distortion, and even quench on the cavity. Therefore, we simply calculated average power densities deposited on the cavity surface with respect to pulse width and pulse repetition rate (PRR). For the calculation, it is assumed that the cavity receives 70 kW from the klystron and the surface area of the 5 cell is 97.96 $cm^2$ outside and 89.36 $cm^2$ inside, respectively. A pulse condition is with less than 1 $W/cm^2$ RF deposition on the outside and inside cavity surface. Although the calculation is based on the assumption that total RF power is deposited on the cavity surface, it appears that the RF pulse condition (1 ms/5 Hz) of 3.9 GHz deflecting mode cavity planned for the ASTA test facility surely exceeds 1 $W/cm^2$, which is about 4 $W/cm^2$ inside and 3.5 $W/cm^2$ outside. We examined the RF-loading characteristics with a full 3D simulation modeling analysis using an integrated multi-physics computational platform.

Simulations show $S_{11}$ spectra of the original and deformed 5 cell structures, obtained from frequency domain solvers before and after RF-loading of 1 ms pulse to the cavity. The displacement of thermal energy deposition causes 7.32 MHz frequency shift, accompanied with 2.82 dB amplitude change and 59 degree phase deviation, on the $TM_{110,\pi}$ mode, which is beyond the FWHM (full-width-half-maximum) bandwidth of the input signal. The designed system has a thermal sensitivity of 1.32 MHz/K, corresponding to 0.51 dB/K and 10.7 degree/K.

## HIGH POWER RF-TEST

The 5 cell cavity in the cryo-vessel, after completely shielded by thermal insulation foams, was tested at room-

(297 K) and LN$_2$-temperature (80 K). The cold cavity parameters were measured twice after 0 day and 20 days long cooling operations to observe temporal variation of thermal stability. Figure 2 shows S$_{11}$ spectra (amplitude and phase) of the warm and cold cavities and thermal variations of a TM$_{110}$-mode. It appears that the resonant frequency is up-shifted with ~ 12.1 MHz by the thermal transition from T = 297.2 K to T = 80 K, corresponding to 55.74 kHz/K, under the condition that cavity frequency varies linearly with temperature. The resonant frequency further rose to 3.899922 GHz after 20 days cooling, corresponding to converged thermal frequency variation of ~ 57 kHz/K, which is close to the simulation result, 64 kHz/K. The return loss remains steady at the initial cooling, while it is gradually increased to ~ - 10 dB, corresponding to 0.0137 dB/K. Even the reduction of coupling level will leave the system fairly operational as 90 % of driving power is still coupled in the system. The instantaneous phase deviation is ~ 0.18 degree/K, but it is also gradually increased up to 0.447 degree/K after 20 days cryo-operation. Therefore, the measurement implies that in the early stage of LN$_2$-operating mode the system can operate with temporal change of < 5 K in terms of amplitude/phase deviation. However, as the vessel loses cooling efficiency with increasing time, acceptable temperature range for the system drops down to < 2 K, which might exceed ambient thermal fluctuation beyond controllable range of the foam-insulated system. For a better thermal management, a cryo-vessel should thus be designed with a vacuum-insulation, which is currently planned with the ASTA 5-cell deflecting mode cavities.

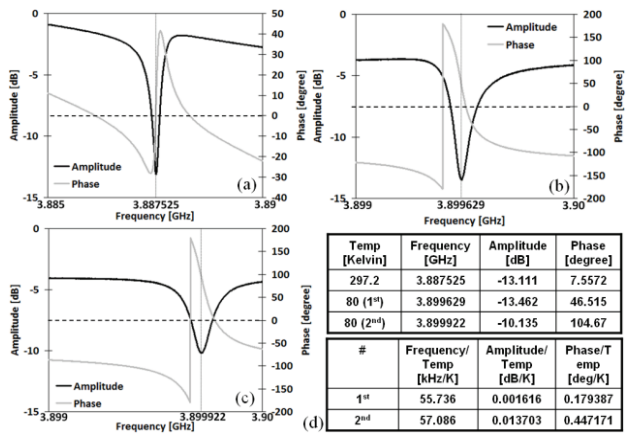

Figure 2: Experimental S$_{11}$-spectal data of the 5 cell test cavity measured at (a) room temperature (T = 297.2 K), and LN$_2$-cooled temperature (T = 80 K) (b) 0-day and (c) 20 days after cooling started.

## BEAM TEST

A deflecting force of the LN$_2$-cooled transverse deflecting cavity (TDC) is tested with a beam of $p$ ~ 3.5 MeV/c and bunch length ($\sigma_z$) ~ 5 ps. We scanned RF launching phases to extrapolate transverse kick strength, $k = eV_\perp/E\rho$. Figure 3 is the phase-scanned beam position and envelope graphs. The drift distance of the beam monitor measuring a peak-to-peak deflection ~ 17 mm is ~ 1.22 m from the TDC, which corresponds to the deflection voltage of ~ 24.4 kV. The kick strength, $k$, from the data is calculated to be ~ 0.41 m$^{-1}$. We will continuously test the cavity to characterize a $R_{65}$ term.

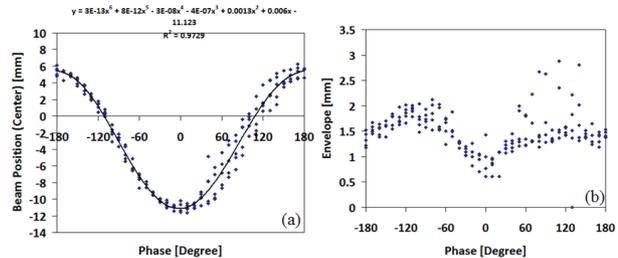

Figure 3: Beam Test result ($p$ ~ 3.5 MeV/c): phase scans of (a) beam center position and (b) envelope.

## CONCLUSION

A normal conducting multi-cell deflecting mode cavity has been used for various beam control applications in an LN$_2$-vessel at the Fermilab A0PI and currently it is planned to accommodate vacuum-cryomodules of either normal conducting or superconducting deflecting mode cavities in the ASTA beam line for higher energy beam tests. Despite successful test results of the previous experiments on 6D phase-space manipulation, the 5-cell cavity demonstrated limited performance. The preliminary high power RF-test and beam test were done and currently systematic optimization of the sensitivity modeling analysis on the deflecting mode vacuum-cryomodule is under consideration in comparison with experimental data.

## ACKNOWLEDGMENT

We would like to acknowledge Philippe R. G. Piot for the support of the experiment. We also thank James Santucci, Jerry M. Makara, and Christopher Prokop for the technical support of the system setup for the experiment.

## REFERENCES

[1] K. Bishofberger et al., Proc. of LINAC 2002, TU404 (2002).
[2] Y.-E Sun et al., Phys. Rev. ST-AB 7 (2004) 123501.
[3] P. Piot et al., Phys. Rev. ST-AB 9 (2006) 031001.
[4] J. Ruan et al., Phys. Rev. Lett. 106 (2011) 244801.
[5] Y.-E Sun et al., Phys. Rev. Lett. 105 (2010) 234801.
[6] T. W. Koeth, "An Observation of a Transverse to Longitudinal Emittance Exchange at the Fermilab A0 Photoinjector," PhD Thesis, 2009.
[7] T. W. Koeth, Proc of PAC2007 Albuquerque, NM, THPAS079, p. 3663 (2007).